\documentclass[12pt]{revtex4}
\def\bea{\begin{eqnarray}}
\def\eea{\end{eqnarray}} 
\def\be{\begin{equation}}
\def\ee{\end{equation}}
\def\nn{\nonumber}

\def\e{\epsilon}

\def\G{\Gamma}

\def\t{\tau}

\begin{document}
\title{Entropic Origin of Hawking Radiation \footnote{write up of talk given at 12th Marcel Grossmann meeting, 2009.}}
\author{Arundhati Dasgupta\\}
\affiliation{Department of Physics and Astronomy,\\University of Lethbridge,\\ 4401 University Drive, Lethbridge T1K 3M4}
\begin{abstract}
We describe a black hole slice by a density matrix and evolve the slice in semiclassical time to search for the origin of Hawking radiation. 
\end{abstract}

\maketitle
\section{Introduction}
Hall, Thiemann, Winkler and collaborators 
derived semiclassical states for canonical variables in loop quantum gravity (LQG) \cite{hall, thiem}. Using the formulation, a `coherent state' for a black hole was defined \cite{adg1}.
This state is sharply peaked on a `classical' non-rotating black hole time slice whose points are sampled by LQG variables defined on a graph with vertices and links. The approximation is good for horizons of at least $10^3 l_p$ or higher. If one traces over the
coherent state inside the horizon, one obtains a density matrix $\rho$. An evaluation of $- \rm Tr (\rho\ln \rho)$ gives the entropy of the
black hole to be proportional to the area of the horizon. This is a first principle derivation of entropy, it is finite and satisfies the Bekenstein-Hawking law. In \cite{adg3}, the density matrix
was evolved in time as observed by a semiclassical observer. The quasi-local energy of Brown and York serves as the Hamiltonian to evolve the horizon from one time slice to the other.
Under a Hermitian flow, the entropy remains unchanged. Allowing for vertices within the horizon to emerge from behind the horizon, one uses a non-Hermitian flow. 
The entropy changes under this time evolution and Hawking radiation emerges as a real effect.
 In this article, we take a simplified U(1) group to illustrate and explain the derivation of entropy and Hawking radiation. {(\it The actual
black hole is described by SU(2) coherent states)}. In the next section we describe the derivation of a density matrix for U(1) group to explain the origin of entropy. In the third section we evolve the system in time to show the origin of emission from the
horizon. 
\section{Density Matrix derived from a U(1) coherent state}
Let us take the simplified case of a U(1) abelian coherent state defined for the variables $h_e,p_e$ which are canonically conjugate Loop Quantum Gravity variables, the $e$ is a 
discrete label. The U(1) coherent state in the momentum representation is given by
\be
\psi^t= \left(\frac{t}{\pi}\right)^{1/4} e^{-p_e^2/2t} e^{- (n^2 t)/2}e^{i n(\chi_e- i p_e)} 
\ee
where $\exp(\iota\chi_e), p_e$ represent the classical holonomy and momentum of the usual loop quantum gravity variables, t controls the width of the distribution 
$n$ is an integer, the momentum quantum number. 
If we take the product of two such states, we get
\be
\psi^t=\sqrt{\frac{t}{\pi}}e^{-p^2_{e_1}/2t} e^{-p^2_{e_2}/2t} e^{- (n^2 t)/2} e^{i n(\chi_{e_1} - i p_{e_1})} e^{- (m^2 t)/2} e^{i m (\chi_{e_2} - i p_{e_2})}
\ee

If the edges $e_1$ and $e_2$ are adjacent, then  $\chi_{e_2}, p_{e_2}$ is given, by the classical eom $\chi_{e_2} = \chi_{e_2}(\chi_{e_1}) , p_{e_2}= p_{e_2}(p_{e_1})$.
(e.g. let us say the field in the continuum satisfies $\frac{d\chi}{dx} =  \chi$, in one dimension x which is then discretised into edges of length $e$. For two adjacent edges, $\chi_{e_2}=\chi_{e_1} e^{e}$).
A density matrix built from the above state will be such that 
\bea
\rho_{nmn'm'} &=& \left(\frac{t}{\pi}\right)e^{- p_{e_1}^2/t - p_{e_2}^2/t}e^{- (n^2 t)/ 2 -i n (\chi_{e_1} + i p_{e_1})} e^{-(m^2 t)/2 -i m (\chi_{e_2} + i p_{e_2})} \nn \\
&& \times \ e^{- (n'^2 t)/2 + i n'(\chi_{e_1} - i p_{e_1})} e^{-(m'^2 t)/2 + i m'(\chi_{e_2} - i p_{e_2})}
\label{den1}
\nn\eea

The reduced density matrix obtained by tracing over the second edges would give us
\be
\bar\rho_{nn'}=\sum_{m} \left(\frac{t}{\pi}\right) e^{- p_{e_1}^2/t}e^{-(n^2 t)/2} e^{-i n (\chi_{e_1}+ i p_{e_1})} e^{-(mt - p_{e_2})^2/t} e^{-(n'^2 t)/2} e^{i n'(\chi_{e_1}- i p_{e_1})} 
\ee
In the $t\rightarrow 0$ limit the sum over m collapses to a delta function in the variable $mt$ and one obtains non-zero elements for the density matrix at $\bar m t = p_{e_2}$. Due to the fact that $p_{e_1}(p_{e_2})= p_{e_1}(\bar mt)$, the density matrix does not factorise.
In the $t\rightarrow 0$ limit, the diagonal elements of the density matrix dominate. 
\be
\bar\rho_{nn} = \delta(nt, p_{e_1}(\bar mt))
\ee

This is precisely the case for the SU(2) coherent state defined in \cite{adg1} for a black hole time slice. Infact equation (123) of the paper shows the density matrix as
\be
\rho_{v_1j_{O}j_H,j_{O}j_H}= |f|^2\delta(j_{O}t,P^I_{O})\delta(m_{O}t,P^I_{O})\delta(j_H t, P_H)\delta(m_Ht, P^I_H) 
\label{den}
\ee
where $|f|^2$ is a normalisation constant. In the above $v_1$ labels a vertex outside the horizon, $j_{O},m_{O}=-j_{O}.....+j_{O},j_H,m_H=-j_H....+j_H$ label the SU(2) angular momentum numbers of the state outside the horizon and at the horizon. The
state inside the horizon has been traced over. The $P^I_{O,H}$ (I=1,2,3) are the classical `momenta' of the LQG phase space. Instead of imposing a conditional probability to ensure the correlations as was stated in \cite{adg2}, the correlations are naturally
encoded in the delta functions. However, the classical data is not enough to uniquely determine the state outside the horizon. The $P^I_H$ takes the $2j_H+1$ values.  
There is thus a degeneracy of $2j_H+1$ at each vertex $v_1$. A calculation of $-\rm Tr (\rho\ln \rho)$ for the product of all vertices surrounding the horizon, gives rise to the
Bekenstein-Hawking entropy. 
\section{Origin of Hawking Radiation}
We evolve the density matrix derived in (\ref{den1}) using a Hamiltonian in the frame of a semiclassical observer stationed just outside the horizon.
\be
i \hbar \frac{\partial \rho}{\partial \t}= [ H, \rho]
\ee

where $H$ is a Hamiltonian. This gives the reduced density matrix at a later slice $\delta \tau$ away from the initial slice at $\tau=0$ to be
$\bar\rho^{\delta \t} = \bar\rho^{0} -\frac{i}{\hbar} \delta \t A$ where $A$ represents the matrix elements of the $[H, \rho]$ commutator in the reduced system. Clearly the entropy in the evolved slice evaluated as $S_{\rm BH}^{\delta \t}= -Tr(\bar\rho \ln\bar \rho)$ can be found as
\be
S_{\rm BH}^{\delta \t} = S_{\rm BH}^{0} +\frac{i}{\hbar} \delta \t [{\rm Tr} A \ln \bar\rho^0 + {\rm Tr}\bar\rho^{0}\bar\rho^{0 \ -1} A]
\label{change}
\ee
 We take the Brown and York quasi local energy at the horizon as the Hamiltonian in the frame of a semiclassical observer.
At the horizon, it has a special form, as one can use the apparent horizon equation. For the U(1) group one can approximate this by $H= \sum_v H_v=\frac{-i}{8\pi G}\sum_v C_v h_{e}^{-1}\beta\frac{\partial h_e}{\partial \beta} P_e +$ Hermitian conjugate.
We have used a derivative of the holonomy w.r.t. Immirzi parameter, which isolates the extrinsic curvature
in the gauge connection of LQG defined as $A_a^I= \G_a^I-\beta K_a^I$ where $\G_a^I$ is the spin connection, and $K_a^I$ is the extrinsic curvature. This procedure of giving the Immirzi parameter a variable status is similar to allowing the dimension to vary in dimensional regularisations. The $C_v$ is a set of constants for each
vertex $v$ immediately outside the horizon. One obtains \cite{adg3} 
$S_{\rm BH}^{\delta \t} - S_{\rm BH}^{0} = 0$. This is expected from classical physics. To obtain the origin of Hawking radiation, one has to allow the vertices within the horizon in one time slice to evolve
to outside the horizon.  The Hamiltonian is a function of derivatives of the normal vector to the constant radii surfaces and within the horizon this has imaginary components as its norm $(1-r_g/r)$ is negative ($r_g$ is the radius of the horizon). For U(1), the Hamiltonian at one vertex is thus $i \e H_v $ where $\epsilon$ is a dimensionless constant. The operator is not Hermitian, and $A= H_v\rho- \rho H^{\dag}_v$ . One gets a real net change in entropy from (\ref{change}). In terms of the black hole parameters, assuming that the U(1) case
simulates the leading terms of the SU(2) calculation, the change in entropy is given by 
\be
- \delta \tau \frac{4\pi r_g \epsilon C}{l_p^2} 
\ee
This is exactly the change in entropy due to emission of a particle of energy given by $G\hbar \omega= \delta \tau \epsilon C$, C is a constant, as $4\pi(4G^2(M- \hbar\omega)^2)/4l_p^2- 4\pi G^2M^2/l_p^2 \approx - 4\pi r_g (G\hbar\omega)/l_p^2$.  


\end{document}